\documentstyle[emulateapj]{article}
\begin{document}

\title{Iron Abundance Profiles of 12 Clusters of Galaxies
Observed With {\it BeppoSAX}}

\author{Jimmy A. Irwin\altaffilmark{1} and Joel N. Bregman}
\affil{Department of Astronomy, University of Michigan, \\
Ann Arbor, MI 48109-1090 \\
E-mail: jirwin@astro.lsa.umich.edu, jbregman@umich.edu}

\altaffiltext{1}{Chandra Fellow.}

\begin{abstract}

We have derived azimuthally-averaged radial iron abundance profiles
of the X-ray gas contained within 12 clusters of galaxies with redshift
$0.03 \le z \le 0.2$ observed with {\it BeppoSAX}.
We find evidence for a negative metal abundance gradient in most of the
clusters, particularly significant in clusters that possess cooling flows.
The composite profile from the 12 clusters resembles that
of cluster simulations of Metzler \& Evrard (1997). This abundance gradient
could be the result of the spatial distribution of gas-losing galaxies
within the cluster being more centrally condensed than the primordial hot gas.
Both inside and outside the core region, we find a higher abundance in
cooling flow clusters than in non-cooling flow clusters. Outside of the cooling
region this difference cannot be the result of more efficient sputtering of
metals into the gaseous phase in cooling flow clusters,
but might be the result of the mixing of low metallicity gas from the outer
regions of the cluster during a merger.

\end{abstract}


\section{Introduction} \label{sec:intro}

The presence of metals in the hot gas contained within clusters of
galaxies provides important clues as to how metals are deposited into the
intracluster medium (ICM) from the constituent galaxies of the cluster.
The two most likely mechanisms for injecting metals into the ICM are
through ram pressure stripping of the gas from galaxies by the ICM
(e.g., Gunn \& Gott 1972; Gaetz, Salpeter, \& Shaviv 1987) and
galactic winds (e.g., Ostriker \& Yahil 1973; Metzler \& Evrard 1994).
Recent work has suggested that negative radial abundance gradients are
a common feature in clusters. Analysis of {\it ASCA} data have indicated
that in some clusters the metal abundance declines from $\sim$40\%--60\%
of solar in the center to $\sim$20\% of
solar at a distance of 0.5--1 Mpc (Ikebe et al.\ 1997; Ezawa et al.\ 1997;
Sarazin, Wise, \& Markevitch 1998; Finoguenov, David, \& Ponman  1999;
Dupke \& White 2000a,b)

Most of the above studies, however, dealt with groups or cooler clusters. The
large, energy-dependent Point Spread Function (PSF) of {\it ASCA} preferentially
scatters hard photons, making spatially-resolved spectroscopy problematic for
hot clusters. Whereas the problem is less severe for cooler ($< 5$ keV)
clusters, substantial errors in the radial temperature and abundance profiles
can occur if the PSF is not dealt with properly (see Takahashi et al.\ 1995).
As such, little observational work has been done on the radial abundance
profiles of hotter clusters. There has also been correspondingly little
theoretical work performed on the predicted abundance profiles of clusters from
hydrodynamical simulations. White (1999) has derived PSF-corrected abundance
profiles for a large sample of clusters observed with {\it ASCA}, but the
deconvolution technique used to correct for the PSF unfortunately led to rather
large uncertainties in the derived abundance profiles. As a result, only weak
constraints could be placed on the presence of abundance gradients from these
data.

{\it BeppoSAX} is better-suited to perform spatially-resolved spectroscopy on
hot clusters. The PSF of {\it BeppoSAX} is one-half that of the {\it ASCA} GIS,
and more importantly, is only weakly dependent on energy. The abundance
profiles of five higher temperature clusters have already been derived from
{\it BeppoSAX}, finding at least marginal evidence for abundance gradients in
A2029, A2256, A3266, and PKS0745-191, with no evidence for a gradient in A2319
(Molendi \& De Grandi 1999; Molendi, De Grandi, \& Fusco-Femiano 2000;
De Grandi \& Molendi 1999a,b; Molendi et al.\ 1999).
In this {\it Paper}, we increase this sample size by
analyzing {\it BeppoSAX} archival data for 12 clusters of galaxies, and derive
radial abundance profiles for each cluster, nine of which have temperatures
greater than 5 keV. Throughout this paper, we assume
$H_0=50$ km s$^{-1}$ Mpc$^{-1}$ and $q_0=0.5$.

\section{Sample and Data Reduction} \label{sec:sample}

From the {\it BeppoSAX} Science Data Center (SDC) archive (available at
http://www.sdc.asi.it/sax\_main.html) we have obtained data for 12 clusters
of galaxies. Within this sample, eight of them possess cooling flows
(A85, A496, A1795, A2029, A2142, A2199, A3562, and 2A0335+096) and four do not
(A2163, A2256, A2319, and A3266). We define a cluster as having a cooling
flow if the cooling rate is greater than $20~M_{\odot}$ yr$^{-1}$;
see Peres et al.\ 1998 and White, Jones, \& Forman  1997. All of the clusters
are at low redshift ($0.03 \le z \le 0.09$) except for A2163, which has a
redshift of $z=0.203$. We analyze data taken with the Medium Energy Concentrator
Spectrometer (MECS) onboard {\it BeppoSAX}. The MECS (see Boella et al.\ 1999
for details) is composed of two
identical gas scintillation proportional counters (three detectors
before 1997 May 9) that are sensitive in the 1.3--10.5 keV energy range.
The event files for all 12 clusters were subjected to the standard
screening criteria of the {\it BeppoSAX} SDC.

Since we are interested in deriving the radial abundance profiles for these
clusters, it is important to account for scattering owing to the
(PSF) of the MECS instrument. Fortunately,
the detector + telescope PSF of {\it BeppoSAX} is nearly independent of energy.
This is because the Gaussian PSF of the MECS
detector improves with increasing energy, while the PSF of the grazing
incidence Mirror Unit degrades with increasing energy
(D'Acri, De Grandi, \& Molendi 1998), leading to a partial cancellation when
these two effects are combined.

To correct for the PSF we have used the routine {\it effarea}, available
as part of the SAXDAS 2.0 suite of {\it BeppoSAX} data reduction programs.
The program {\it effarea} convolves the surface brightness profile of
the cluster (which has been determined using {\it ROSAT} PSPC data;
see Mohr, Mathiesen, \& Evrard 1999 and Ettori \& Fabian 1999) with the
PSF of the MECS to determine the amount of contamination in the spatial region
in question at each energy from other regions of the cluster. Thus, an
energy-dependent correction vector is formed.
When this correction vector is multiplied by the observed spectrum, the
PSF-corrected energy spectrum is obtained for the region. In practice, this
information is incorporated into the auxiliary response file (the {\it .arf}
file), which is subsequently used in the spectral fitting. This task also
corrects for vignetting. A more complete description of the task is given in
Molendi (1998) and D'Acri et al.\ (1998). This correction does not appear
to affect the spectrum greatly; D'Acri et al.\ (1998)
and Kaastra, Bleeker, \& Mewe (1998) found only small changes between the
uncorrected and corrected temperature profiles for Virgo and A2199,
respectively. In addition, D'Acri et al.\ (1998) found that the correction
vectors amounted to 5\% or less for energies above 3 keV outside of the
innermost bin (the innermost bin loses some flux via scattering but gains
very little from photons scattered in from greater off-axis radii).
Our inner three spatial bins are identical in angular extent to the bins
of D'Acri et al.\ (1998), so we assume that our correction vectors are
as small as theirs.
In addition, we find only a small difference between our corrected and
uncorrected abundance profiles for our sample.

We extracted spectra in four annular regions for each cluster, with inner
and outer radii of $0^{\prime}-2^{\prime}$, $2^{\prime}-4^{\prime}$,
$4^{\prime}-6^{\prime}$, and $6^{\prime}-9^{\prime}$. At $9^{\prime}$ the
telescope entrance window support structure (the strongback) strongly absorbs
X-rays, making spectroscopy difficult in this area of the detector.
In addition, for off-axis angles greater than $10^{\prime}$, the departure
of the PSF from radial symmetry becomes noticeable (Boella et al.\ 1997). With
this in mind, we ended the radial profiles at $9^{\prime}$. At this radius,
the profiles extended to 17\%--33\% of the virial radius, where
$r_{virial} = 3.9~(T/10~{\rm keV})^{1/2}$ Mpc (see Evrard, Metzler, \& Navarro
1996), for all clusters except A2163, for which the profile extended to 55\% of
$r_{virial}$. We also extracted one global
spectrum ($0^{\prime}-9^{\prime}$) and also two spectra covering
the two regions $0-0.075r_{virial}$ and
$0.075-0.173r_{virial}$. Background was obtained from the deep blank sky
data provided by the SDC. We used the same region filter to extract the
background as we did the data, so that both background and data were affected
by the detector response in the same manner. The energy channels were
regrouped to contain at least 25 counts.

For each cluster, XSPEC Version 11.0 was used to fit the spectrum. The MECS2
and MECS3 (and MECS1 when available) data were fit separately, but with the
same temperature, metallicity, and normalization. We assumed a MEKAL model
with an absorption component fixed at the Galactic value, allowing the
temperature and metallicity to vary. For the iron-to-hydrogen ratio we use a
value of $4.68 \times 10^{-5}$ (Anders \& Grevesse 1989) since this is the value
assumed for most previous abundance determinations, although it should be noted
that more recent determinations of the Fe/H ratio point to a value of
$3.24 \times 10^{-5}$ (Ishimaru \& Arimoto 1997). Because of a systematic
shift of 45--50 eV at 6.6 keV in the MECS channel--to--energy conversion,
we have also allowed the redshift to vary (F. Fiore, private communication).
This systematic shift was
evident in our sample; when the redshift was allowed to vary, the measured
redshift was less than the optically-determined redshift in all 12 clusters,
and inconsistent with the optically-determined redshift at the 90\% confidence
level for eight of them. A modest decrease in the reduced $\chi^2$ also
occurred for most of the clusters when the redshift was allowed to vary.
However, freeing the redshift did not affect the values obtained for the
temperature and metallicity by more than 5\% (and in most instances much less),
so the results do not depend significantly on this shift. We also fit only the
3.0--10.5 keV part of the spectrum because of excess flux found below 3.0 keV
that might be the result of uncertainties in the calibration of the MECS
instruments; see Irwin \& Bregman (2000) for a more complete description.
All quoted errors are 90\% confidence levels for one interesting parameter
($\Delta \chi^2 = 2.71$) unless otherwise noted.
Below we present the results from the abundance analysis. The results of
the temperature analysis are given in Irwin \& Bregman (2000). One additional
cluster (A3562) not analyzed in Irwin \& Bregman (2000) is presented here.

\section{Abundance of the Hot Gas} \label{sec:abun}

\subsection{Global Abundances} \label{ssec:abun_global}

For hot clusters it is the Fe-K$\alpha$ line that primarily determines
the metal abundance measurement. The fits to the global spectra were adequate,
with $\chi_{\nu}^2 < 1.20$ in all cases. For the eight clusters harboring
cooling flows, the weighted average emission-weighted global abundance
was $0.37\pm0.04$ (the error is given by the weighted standard deviation for
the subsample). The four clusters lacking cooling flows had an average
emission-weighted global abundance of $0.27\pm0.02$. The ratio of cooling flow
to non-cooling  flow global abundance is $1.4\pm0.2$, in good agreement with the
value of 1.6 obtained by Allen \& Fabian (1998) with a sample of 30 clusters
observed with {\it ASCA} when they used a single component thermal model.

\subsection{Radial Abundance Profiles} \label{ssec:abun_profiles}

\begin{figure*}[htb]
\vskip4.6truein
\includegraphics{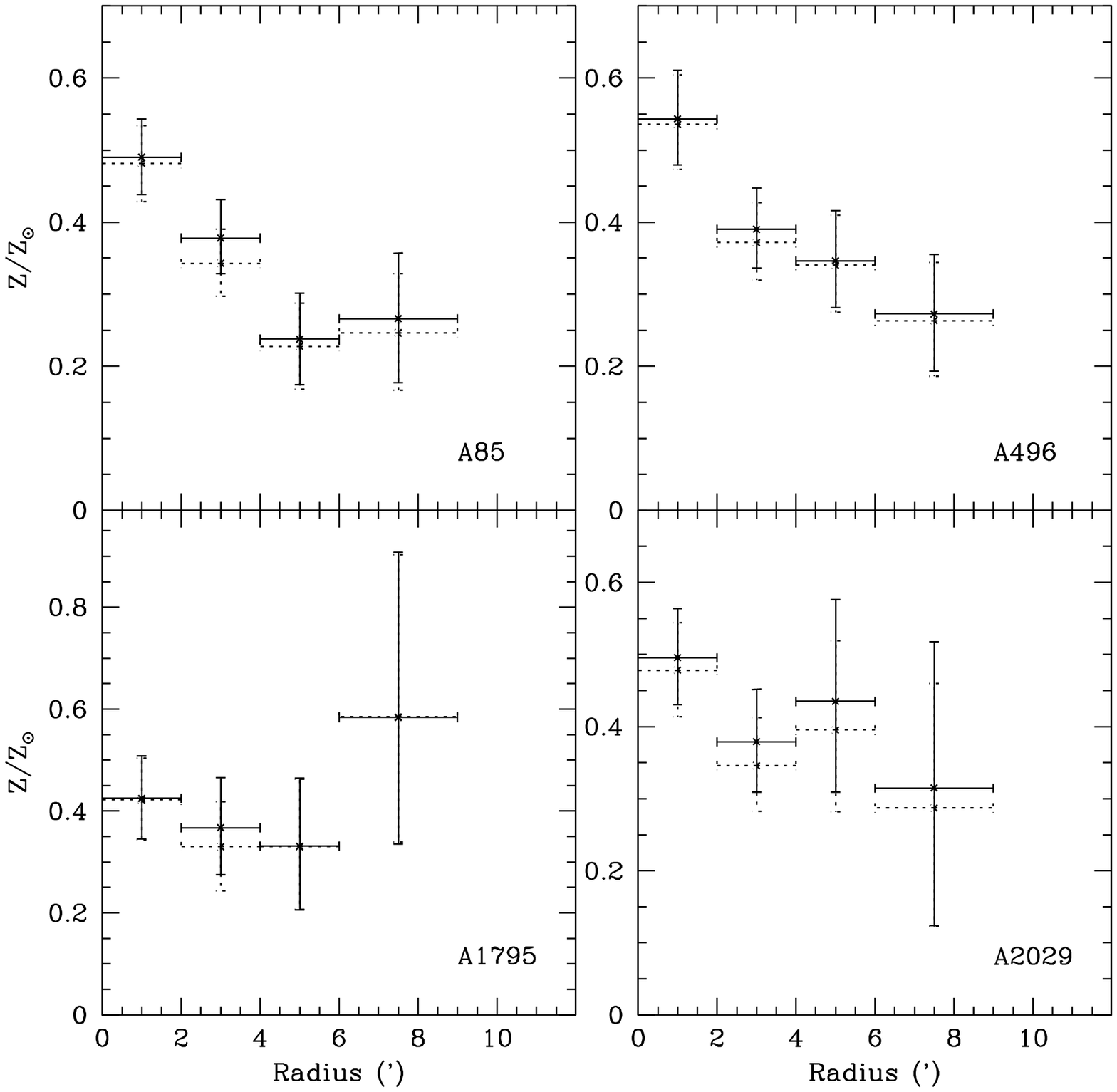}
\caption[abun]{
Radial abundance profiles for the 12 clusters in the sample with 90\%
confidence levels, derived from
spectral fitting in the 3.0--10.5 keV energy range. Solid lines represent
abundances corrected for the PSF of {\it BeppoSAX} and the dotted lines are
uncorrected for the PSF, except for A2256 where the dotted line represents
the {\it BeppoSAX} profile derived by Molendi et al.\ (2000) from a longer
observation (their error bars have been multiplied by 1.65 to convert 68\%
confidence levels to 90\% confidence levels).
\label{fig:abun}}
\end{figure*}

The radial metal abundance profile for each cluster is shown in
Figure~\ref{fig:abun}. Four of the clusters have had abundance gradients
determined with {\it ASCA}, and are generally consistent within the
errors with the profiles determined here: A496 (Dupke \& White 2000b;
Finoguenov et al.\ 1999), A2199 (Finoguenov et al.\ 1999), A2029
(Sarazin et al.\ 1998; Finoguenov et al.\ 1999) and 2A0335+096
(Kikuchi et al.\ 1999). In addition, analysis of this same
{\it BeppoSAX} data for A2029, A2256, A2319, and A3266 by other authors
are in agreement with our results (Molendi \& De Grandi 1999;
Molendi et al.\ 1999; De Grandi \& Molendi 1999), with the exception of A2256
for which Molendi et al. (2000) found a significantly decreasing profile
whereas ours is flat. However, our inability to find an abundance gradient
is the result of the short observing time of the data we used; the total
observing time of the Molendi et al. (2000) observation was 2.6 times longer
than ours, and they were able to place much tighter constraints on the
abundance than we were. In both studies, the 1$\sigma$ error bars overlapped
one another.

In general, the abundance profiles decline with radius. Such a trend
was also found in clusters observed by {\it ASCA} (Finoguenov et al.\ 1999;
Kikuchi et al.\ 1999; Dupke \& White 2000a), although there are notable
exceptions to this trend such as A1060 (Tamura et al.\ 1996). These previous
studies focused primarily on cooler clusters ($kT < 5$ keV). This study
confirms the ubiquity of negative abundance gradients in hotter clusters
(nine of our 12 clusters have $kT > 5$ keV),
as illustrated in Figure~\ref{fig:abun-mean}, where all 12 abundance profiles,
normalized to the global abundance for each cluster, are plotted versus
radius in units of the virial radius. A strong negative gradient in the
abundance profile is evident out to 25\% of the virial radius. However, there
are a few points quite discordant with this trend at large radii. The clusters
responsible for these discordant points are A1795, A2142, and A2163.

\begin{figure*}[htb]
\vskip4.5truein
\includegraphics{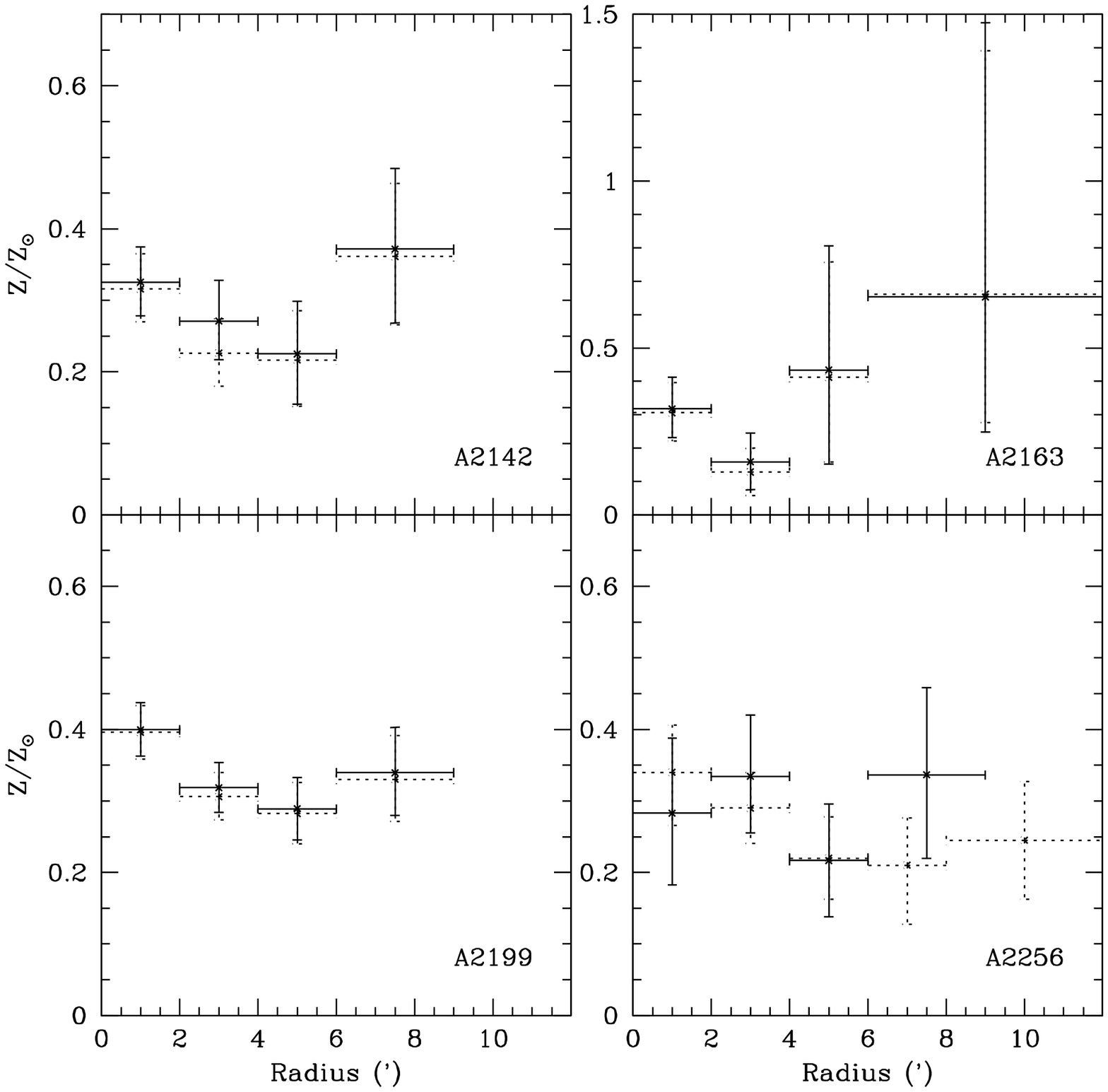}
{
~~~~~~~~~~~~~~~~~~~~~~~~~~~~~~~~~~~~~~~~~~~~~~~~~~~~~~~~~~~~~~~~~~Fig. 1 -- continued.
}
\end{figure*}

\begin{figure*}[htb]
\vskip4.64truein
\includegraphics{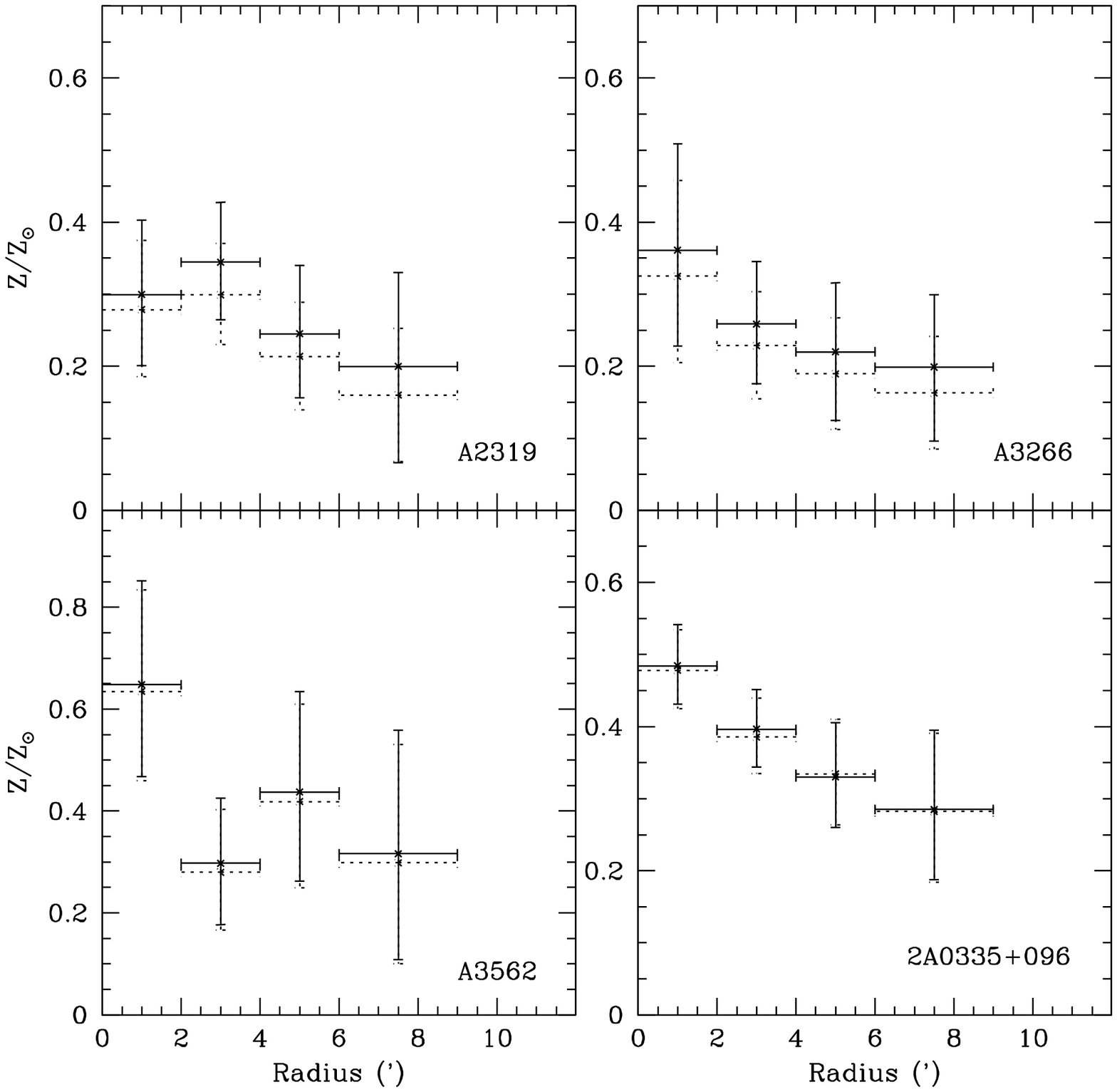}
{
~~~~~~~~~~~~~~~~~~~~~~~~~~~~~~~~~~~~~~~~~~~~~~~~~~~~~~~~~~~~~~~~~~Fig. 1 -- continued.
}
\end{figure*}

We investigated the possibility that the use of single component temperature
models in the central regions of cooling flow clusters has biased the
determination of the abundance in these regions. The addition of a cooling
flow component did not lead to a significant improvement in the fits, and 
the cooling rate was only weakly constrained. This is a result of excluding
data below 3.0 keV, where the effect of cooling flows on the spectra is
most pronounced. When we fixed the cooling rate at values determined by other
satellites, the best-fit abundance value was actually slightly {\it higher}
than in the single component case by a few percent, indicating that the central
enhancement in the abundance is real.

We have fit a linear model of the form $(Z/Z_{mean})=a-b(r/r_{virial})$ to the
data. While including all values, the fit is not very good, with $\chi^2=72.2$
for 47 degrees of freedom, and can be excluded at $>98\%$ confidence level.
Removal of the four discordant data points yielded
an excellent fit, with $\chi^2=46.1$ for 43 degrees of freedom for values
of $a=1.212$ and $b=2.705$ (2.149--3.250 at 90\% confidence). Although it is
not clear why four of the clusters show a  $\sim$2$\sigma$ departure
from the observed trend outside
$\sim$25\% of the virial radius, the derived slope is nonetheless an accurate
representation of the data inside of 25\% of the virial radius. Excluding
non-cooling flow clusters in addition to the discordant points led to the
same best-fit slope.

\subsection{Comparison To Theoretical Models} \label{ssec:theory}

Surprisingly little theoretical work has been performed to date to quantify
the behavior of abundance gradients in clusters. One such study is that
of Metzler \& Evrard (1997), who used gas dynamical simulations which included
the effects of galactic winds to generate an ensemble of 18 realizations
of clusters spanning a wide range of temperatures. Assuming that the iron
present in the gas resulted from the ejection of metals from galaxies, a
composite iron abundance profile for the 18 realizations was determined. It
was found to decrease significantly with increasing radius. This resulted
from the spatial distribution of the galaxies being more centrally condensed
than the original primordial X-ray gas. Since the ejected gas traces the
galaxies, an abundance gradient is created relative to the primordial gas.

\begin{figure*}[htb]
\vskip4.6truein
\includegraphics{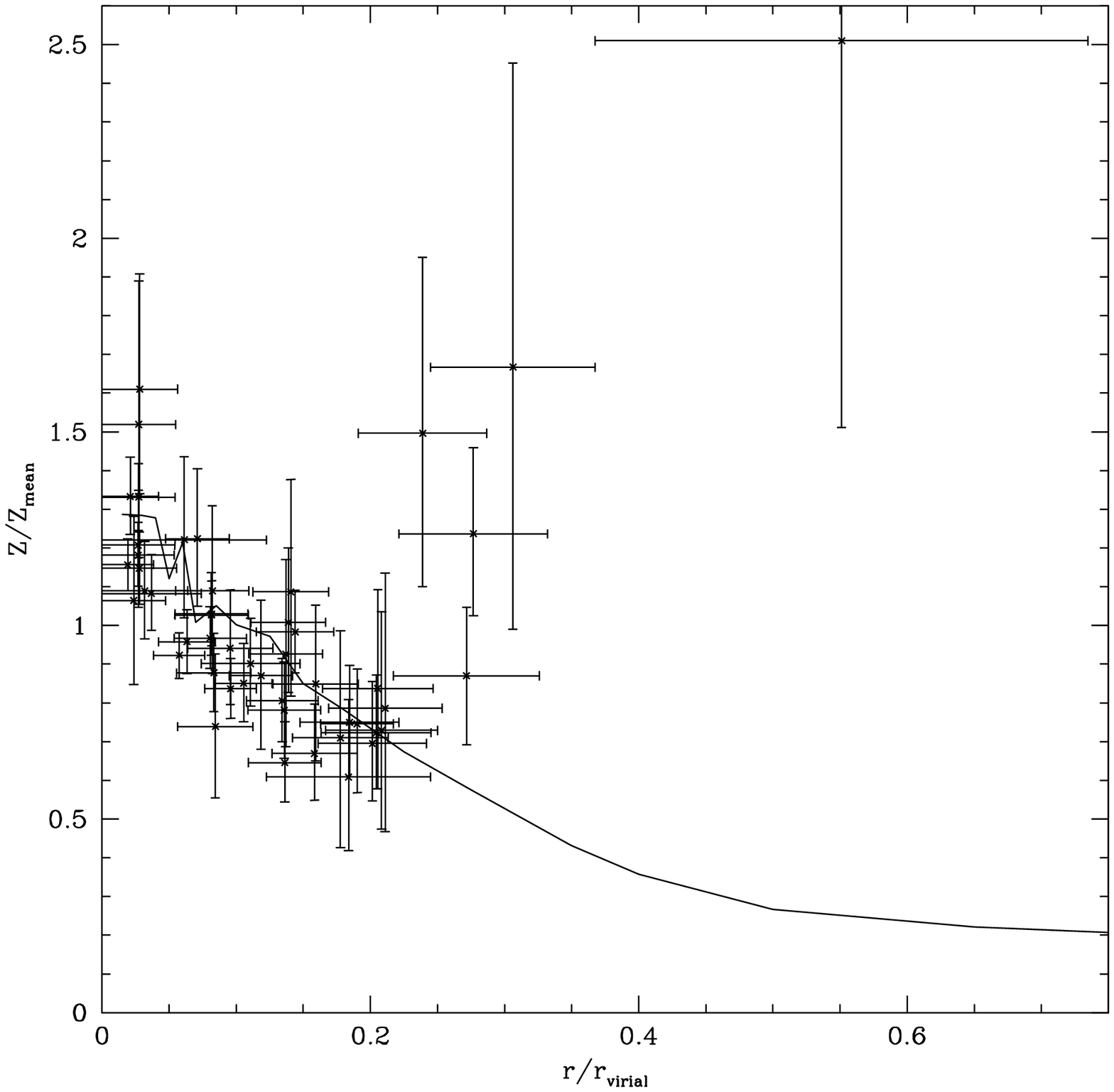}
\caption[[abun-mean]{
Normalized abundance profiles for all 12 clusters in the sample versus
radius in units of the virial radius. Here, the error bars represent the
1$\sigma$ errors. A significant radial gradient in the abundance is evident.
The solid line represents the composite abundance profile of 18 realizations
of the cluster simulations of Metzler \& Evrard (1997).
\label{fig:abun-mean}}
\end{figure*}

The composite iron abundance profile from the 18 three-dimensional realizations
of Metzler \& Evrard (1997) is shown superposed on the {\it BeppoSAX}
data in Figure~\ref{fig:abun-mean} as a solid line. The composite profile
was normalized assuming an average global abundance of 40\% of the solar
value. Metzler \& Evrard (1997) scaled each cluster in radial units of
$r_{170}$ before merging the profiles, where $r_{170}$ is the radius within
which the mean density is 170 times the critical density. This is very close
to our adopted value of $r_{virial}$. The normalized abundance profile of
Metzler \& Evrard (1997) matches the data very well, ranging from a value
of 1.3 in the center to about 0.6 at 25\% of the virial radius.

We note that recent work by Finoguenov et al.\ (1999) and Dupke \& White (2000b)
has shown that significant radial gradients exist in the elemental
abundance ratios in clusters. This would argue against the idea that a
difference in the spatial distribution between the galaxies and the gas is
responsible for the observed abundance gradient if it is assumed that both
Type Ia and Type II elements were released into the ICM under the same
conditions. The observational evidence implies that Type Ia and Type II
elemental contamination is not the same in the inner and outer parts, with
Type Ia ejecta being dominant in the center. Dupke \& White (2000b) propose
that the observed abundance ratios can be explained by a phase of
early, vigorous protogalactic wind driven by Type II supernovae that spread
metals uniformly throughout the cluster, followed by a phase of less vigorous
Type Ia supernovae wind that was suppressed by the central dominant galaxy
in the center of the cluster. Thus, the measured iron abundances originate
from a combination of Type II supernovae (having no radial gradient) and
Type Ia supernovae (having a radial gradient). The gradient we measure is the
linear combination of these two effects.

\subsection{Cooling Versus Non-cooling Flow Clusters} \label{ssec:cool_nocool}

Inspection of the abundance profiles of the eight cooling flow clusters reveals
a radially decreasing trend in all of them. The gradient is quite significant
in A85, A496, A2199, A3562, and 2A0335+096, and less significant
in A1795, A2029, and A2142.
This seems to be a universal trend in cooling flow clusters since all
nine cooling flow clusters in the sample of Finoguenov et al.\ (1999), as well
as all three clusters in the samples of Kikuchi et al.\ (1999) and
Dupke \& White (2000a) show the same trend, at least marginally.
No clear trend exists among the non-cooling flow clusters. The profile of
A2256 is flat, although a decline with radius cannot be ruled out owing to
the large errors (a decline was seen with a longer {\it BeppoSAX} observation
by Molendi et al.\ 2000; see Figure~\ref{fig:abun}).
A2319, A3266, and A2163 show some evidence for a
negative gradient although the profiles are formally consistent with a flat
profile. Results with {\it ASCA} for clusters with little or no cooling flows
are also mixed. Whereas Coma, A401, A1060, and A2670 show a flat abundance
profile, A399 has an abundance gradient (Watanabe et al.\ 1997;
Fujita et al.\ 1996; Tamura et al.\ 1996; Finoguonov et al.\ 1999).
It is possible that mergers that could be responsible for the
disruption of the cooling flow in clusters can also destroy abundance
gradients. The degree to which the abundance gradient remains intact may
be an indicator of how recent and/or violent the merger was. All four
non-cooling flow clusters are merger candidates. Of the five cooling flow
clusters with significant gradients, only A85 is a merger candidate.
Of the three cooling flow clusters with only weak evidence
for an abundance gradients two are recent merger candidates.
Oegerle \& Hill (1994) conclude that the velocity distribution of galaxies
within A1795 suggest that the cluster underwent a merger with a relatively
small subcluster such that the cooling flow was not disrupted.
Oegerle, Hill, \& Fitchett (1995) also find evidence for subclustering
from the galaxy velocity distribution in A2142. {\it ROSAT} PSPC and HRI
images of A2142 also suggest that a merger is in process (Henry \& Briel 1996).
It is possible that mergers have mixed the gas of these two clusters
enough to decrease the significance of an abundance gradient without destroying
the cooling flow.
A2029 does not show evidence for a merger, although the lack of a
significant abundance gradient may be the result of poor photon statistics
rather than a lack of a gradient. Further studies of A2029 will be needed
to confirm this.

As suggested by Fabian et al.\ (1994) and Allen \& Fabian (1998),
the presence of abundance gradients
in cooling flow clusters provides a reasonable explanation as to why cooling
flow clusters have higher emission-weighted metal abundances than their
non-cooling flow counterparts. If the cooling gas at the center of clusters
has a higher abundance than the outer regions, the strongly peaked X-ray
surface brightness distribution characteristic of cooling flow clusters will
preferentially weight the global emission-weighted abundance towards the central
value, with the low-abundance outer regions contributing relatively little
to the global
value. Such an effect is not present in non-cooling flow clusters owing to the
broader, less-peaked X-ray surface brightness distribution of these clusters.
But is this the only reason for the higher measured metallicity in cooling
flow clusters, or is there really an abundance difference between cooling flow
and non-cooling flow clusters? To test this, we determined abundances in two
regions for each cluster, a circle with a radius of $0.075r_{virial}$
and an annulus with inner and outer radii of $0.075r_{virial}$ and
0.173$r_{virial}$. The outer radius was set by the need to extend the
annulus no further than $9^{\prime}$ for any cluster, and the inner radius
was set to be larger than the cooling radius for any of the cooling
flow clusters.

The results are shown in Table~\ref{tab:abun}. Inside of $0.075r_{virial}$ the
cooling flow subsample had an average abundance of $0.42\pm0.06$,
while the non-cooling flow subsample had an average of $0.33\pm0.04$.
In the outer region, the cooling flow subsample had an average abundance of
$0.30\pm0.02$, while the non-cooling flow subsample had an average of
$0.24\pm0.03$. Thus, there appears to be an abundance gradient in non-cooling
flow clusters as well as cooling flow clusters. The presence of abundance
gradients in non-cooling flow clusters (especially in hot clusters) would seem
to rule out claims that previously observed abundance gradients were artifacts
of incorrect spectral modeling due to either the presence of cooling gas at the
center of clusters or uncertainties in modeling the Fe-L line complex for
cooler clusters.

\begin{table*}[htb]
\caption[Abundances]{}
\label{tab:abun}
\begin{center}
\begin{tabular}{ccccccc}
\multicolumn{7}{c}{\sc Abundances} \cr
\tableline \tableline
&& $kT$ (keV)\tablenotemark{a} & Global & $\chi^2_{\nu}$/d.o.f.&
0--0.075 $r_{virial}$ & 0.075--0.173 $r_{virial}$ \\
\tableline
A85\tablenotemark{\star} && 6.4$^{+0.3}_{-0.2}$ &0.37$^{+0.03}_{-0.03}$
&1.17/282& 0.45$^{+0.04}_{-0.04}$ &0.30$^{+0.04}_{-0.05}$\\
A496\tablenotemark{\star} && 4.2$^{+0.1}_{-0.1}$ &0.41$^{+0.03}_{-0.03}$&
1.20/259 & 0.48$^{+0.05}_{-0.04}$ &0.31$^{+0.05}_{-0.05}$\\
A1795\tablenotemark{\star}&& 6.0$^{+0.4}_{-0.4}$ &0.39$^{+0.05}_{-0.05}$&
0.97/211 & 0.41$^{+0.08}_{-0.07}$ &0.36$^{+0.09}_{-0.09}$\\
A2029\tablenotemark{\star}&& 7.6$^{+0.5}_{-0.4}$ &0.43$^{+0.05}_{-0.04}$&
0.98/141& 0.49$^{+0.06}_{-0.06}$ &0.30$^{+0.08}_{-0.07}$\\
A2142\tablenotemark{\star}&& 8.7$^{+0.4}_{-0.4}$ &0.30$^{+0.03}_{-0.03}$&
1.02/292 & 0.32$^{+0.05}_{-0.05}$ &0.26$^{+0.05}_{-0.05}$\\
A2163&& 11.7$^{+1.0}_{-0.9}$ &0.26$^{+0.06}_{-0.06}$&
1.05/440 & 0.39$^{+0.17}_{-0.14}$ &0.21$^{+0.07}_{-0.07}$\\
A2199\tablenotemark{\star}&& 4.4$^{+0.1}_{-0.1}$ &0.35$^{+0.02}_{-0.02}$&
1.03/419& 0.37$^{+0.03}_{-0.03}$ &0.31$^{+0.04}_{-0.03}$\\
A2256&& 7.1$^{+0.5}_{-0.4}$ &0.28$^{+0.04}_{-0.04}$&
1.06/238 & 0.31$^{+0.09}_{-0.08}$ &0.28$^{+0.06}_{-0.06}$\\
A2319&& 10.5$^{+0.8}_{-0.7}$ &0.28$^{+0.05}_{-0.05}$&
1.04/273 & 0.36$^{+0.07}_{-0.07}$ &0.22$^{+0.06}_{-0.06}$\\
A3266&& 9.9$^{+0.8}_{-0.7}$ &0.24$^{+0.05}_{-0.05}$&
0.93/265 & 0.28$^{+0.10}_{-0.10}$ &0.23$^{+0.07}_{-0.06}$\\
A3562\tablenotemark{\star}&& 5.1$^{+0.6}_{-0.5}$ &0.40$^{+0.08}_{-0.08}$
&0.95/165 &0.53$^{+0.15}_{-0.14}$& 0.30$^{+0.11}_{-0.10}$ \\
2A0335+096\tablenotemark{\star}&& 3.2$^{+0.08}_{-0.08}$ &0.41$^{+0.03}_{-0.03}$
&1.13/257&0.47$^{+0.05}_{-0.05}$ &0.33$^{+0.05}_{-0.04}$ \\
\tableline
\end{tabular}
\end{center}
\tablenotetext{\star}{cooling flow cluster}
\tablenotetext{a}{Best-fit global temperature of {\it BeppoSAX} data from
Irwin \& Bregman (2000), except for A3562 which is presented here for the
first time.}
\end{table*}

Although the average abundances of the cooling flow and
non-cooling flow clusters are marginally consistent within the
errors within the cooling region, the data suggests that there
is a discrepancy between the abundance of cooling and non-cooling flow
clusters outside of the cooling radius. Of course, any conclusions
drawn from a subsample of four clusters should be viewed with caution. A larger
sample of clusters will be necessary to confirm if the abundance differences
between cooling flow and non-cooling flow clusters are real.
This result is in the opposite sense expected if higher metallicity gas
from the cooling flow region has been mixed in with the lower metallicity
gas of the outer regions as the result of a merger in a present-day
non-cooling flow cluster.
This trend also appeared in a sample of clusters observed by {\it ASCA}.
Fukazawa et al.\ (1998) analyzed 26 clusters with temperatures of at least
3.0 keV, and found global abundances after having excised the inner cooling
regions. The average abundance for the 16 cooling flow clusters
was $0.29\pm0.07$, while the average for the ten non-cooling flow
clusters was $0.23\pm0.05$. Although at a lower significance than our
{\it BeppoSAX} sample, the trend is in the same direction.
Fukazawa et al.\ (1998) did not claim to extract their spectra in units
of the virial radius, which is the unit of distance that allows the most
direct comparison of radial properties for clusters of varying
sizes and temperatures. This might wash out some of the information
on the abundance gradient when the profiles are combined, and lead to
a result of lower statistical significance.

Allen \& Fabian (1998) present the interesting scenario in which metallicity
gradients can be explained (at least in part) by a significant fraction of
the metals in clusters residing on grains in cluster cores. Since the
lifetime of a grain to sputtering in the hot ICM is inversely proportional
to the electron density of the hot gas (e.g., Draine \& Salpeter 1979), more
metals would be released into the gas phase in the high density cooling flow
regions of clusters than in lower density regions, providing the dust grains
are $\ge 10$ microns. By extending this argument outside the core region, it
might be possible to explain the difference in metallicity at large radii
between cooling flow and non-cooling flow clusters. If the density of gas
outside the
core is greater in cooling flow clusters than in non-cooling flow clusters,
more metals will be released into the gas phase for the former. A difference
in density among cooling and non-cooling flow clusters outside of 200 kpc
was observed by White et al.\ (1997), who deprojected the
surface brightness profiles of 207 clusters observed with {\it Einstein
Observatory} and found that the electron densities of clusters with cooling
rates greater than $10~M_{\odot}$ yr$^{-1}$ were larger than for clusters
with cooling rates less than $10~M_{\odot}$ yr$^{-1}$ out to a radius of 1 Mpc.
For a 7 keV cluster, our $0.075-0.173$$r_{virial}$ bin corresponds to
240--560 kpc. At a radius of 450 kpc, White et al.\ (1997) found that the
cooling flow clusters had an electron density of $\sim$$7 \times 10^{-4}$
cm$^{-3}$, whereas the non-cooling flow clusters had an electron
density of $\sim$$3 \times 10^{-4}$ cm$^{-3}$. For a grain lifetime of
$2 \times 10^6$ $a/n_e$ yrs (where $a$ is the size of the grain in microns
and $n_e$ is the electron density of the hot gas in cm$^{-3}$), grains with a
size of around 2 microns would survive in non-cooling flow clusters for a
Hubble time while being sputtered in cooling flow clusters. While rather large,
dust grains of this size are not unreasonable. In addition, this argument
assumes that the dust was deposited into the ICM at the time the cluster was
created. In reality, dust is deposited into the ICM throughout the history
of the cluster, allowing smaller dust grains to survive in low density
non-cooling flow clusters while still being sputtered in high density
coling flow clusters. This in turn can lead to a higher abundance
in the outer regions of cooling flow clusters since more of the grains
have been sputtered away owing to the higher electron density.

However, more recent determinations of the density profiles of clusters
determined with {\it ROSAT} PSPC data indicate that the non-cooling flow
clusters in our sample do not have a lower electron density in the
0.075--0.173 $r_{virial}$ region than the cooling flow clusters (J. Mohr,
private communication). At 10\% of the virial radius, the average electron
density for the non-cooling flow clusters A2256, A2319, and A3266 was
$1.80 \pm 0.20 \times 10^{-3}$ cm$^{-3}$, while for the cooling flow clusters
A85, A496, A1795, A2029, A2142, A2199, and A3562 the average electron was
$1.92 \pm 0.46 \times 10^{-3}$ cm$^{-3}$
Thus, it does not appear likely that the difference in abundances outside
the cooling radius can be the result of an increased sputtering of grains
in cooling flow clusters.

Since non-cooling flow clusters are usually merger candidates,
one possible explanation for the difference in abundances between cooling
and non-cooling flow clusters is the mixing of gas via mergers from larger
($> 0.175 r_{virial}$) radii rather than from gas interior to this region.
If the iron abundance of the gas continues to decline with increasing
radius beyond $0.2 r_{virial}$ as the simulations of Metzler \& Evrard (1997)
imply, the mean metallicity in our 0.075--0.173 $r_{virial}$ bin will be
lowered upon mixing with this low metallicity gas during a merger. A2142,
a cooling flow cluster that is thought to be undergoing a merger and shows
only weak evidence for an abundance gradient, also has the lowest metallicity
in the 0.075--0.173 $r_{virial}$ bin among cooling flow clusters in our sample,
consistent with the merger-mixing theory. However, A1795, which is also
believed to have recently experienced a merger has a high metallicity value
in the 0.075--0.173 $r_{virial}$ bin, although the uncertainty is rather
large. A larger sample of clusters with abundances determined out to larger
radii will be needed to explore this topic further.

\section{Conclusions} \label{sec:conclusions}

We have analyzed 12 clusters with the MECS instrument onboard {\it BeppoSAX}
and find evidence for a negative abundance gradient in most of them. This
gradient is present in all eight cooling flow clusters, and to a lesser
significance in the non-cooling flow clusters. Since $kT > 5$ keV for
nine of the 12 clusters, this work extends the ubiquity of abundance
gradients to hot clusters. The slope of the
abundance gradient is in good agreement with cluster simulations performed
by Metzler \& Evrard (1997). In our sample
cooling flow clusters had a higher metallicity than non-cooling flow clusters
both in the inner regions of the cluster and the outer regions out to several
hundred kpc; further work with a larger sample will be needed to confirm
this result for clusters in general. Outside the cooling region, it seems
unlikely that this difference can be explained by the more efficient sputtering
of dust grains in cooling flow clusters, since the densities of cooling
flow and non-cooling flow clusters are similar in this region. The mixing of
low metallicity gas during a merger may be responsible for this difference.

\acknowledgments
We thank the anonymous referee for many useful suggestions and comments.
We thank Chris Metzler for making his theoretical abundance profiles
available to us. We also thank Joe Mohr for kindly providing us his
{\it ROSAT} PSPC density profiles. 
JAI thanks Renato Dupke and Gus Evrard for many useful comments and
conversations.
This research has made use of data obtained through the {\it BeppoSAX}
Science Data Center and the High Energy
Astrophysics Science Archive Research Center Online Service,
provided by the NASA/Goddard Space Flight Center.
This work has been supported by {\it Chandra} Fellowship grant PF9-10009,
awarded through the {\it Chandra} Science Center. The {\it Chandra} Science
Center is operated by the Smithsonian Astrophysical Observatory for NASA
under contract NAS8-39073.

\end{document}